# Single-stage few-cycle nonlinear compression of milliJoule energy Ti:Sa femtosecond pulses in a multipass cell

Louis Daniault[1,*], Zhao Cheng[1], Jaismeen Kaur[1], Jean-François Hergott[2], Fabrice Réau[2], Olivier Tcherbakoff[2], Nour Daher[3], Xavier Délen[3], Marc Hanna[3], Rodrigo Lopez-Martens[1]

[1]*Laboratoire d'Optique Appliquée, CNRS, Ecole Polytechnique, ENSTA Paris, Institut Polytechnique de Paris, 181 chemin de la Hunière et des Joncherettes, 91120, Palaiseau, France*
[2]*Université Paris-Saclay, CEA, CNRS, LIDYL, 91191, Gif-sur-Yvette, France*
[3]*Université Paris-Saclay, Institut d'Optique Graduate School, CNRS, Laboratoire Charles Fabry, 91127, Palaiseau, France*
*\*Corresponding author: louis.daniault@ensta-paris.fr*

**We report on the nonlinear temporal compression of mJ energy pulses from a Ti:Sa chirped pulse amplifier system in a multipass cell filled with argon. The pulses are compressed from 30 fs down to 5.3 fs, corresponding to two optical cycles. The post-compressed beam exhibits excellent spatial quality and homogeneity. These results pave the way to robust and energy-scalable compression of Ti:Sa pulses down to the few-cycle regime.**

## 1. INTRODUCTION

Nonlinear post-compression of ultrashort laser pulses through self-phase modulation (SPM) and dispersion compensation is now widely used to generate pulses in the few-cycle regime for a number of applications [1]. For pulse energies comprised between a few 100 µJ and a few mJ, this is commonly performed via nonlinear interaction with noble gases in a hollow-core fiber (HCF) [2,3], which provides spatially homogeneous spectral broadening over extended nonlinear propagation lengths. Its main limitation comes from ionization that arises at high intensities. Higher input energies imply larger capillary cores with correspondingly longer interaction lengths, which can quickly become impractical for table-top use.

For the past few years, multipass cells (MPC) have been employed for compressing ultrashort pulses as a promising alternative to HCF [4,5]. The nonlinear interaction takes place in free-space propagation, is distributed over much larger path lengths than in capillaries and provides a large spectral broadening with reduced spatio-temporal couplings. Recent experiments report compression factors comparable to those obtained with HCF along with excellent efficiency [5-10]. To date, MPCs have been almost exclusively dedicated to Yb laser-based nonlinear compression experiments, which often require a double-stage compression scheme in order to reach the few-cycle regime. From pulse ranging from several 100 fs up to 1 ps duration, a first stage involves dielectric mirrors with high damage threshold and high reflectivity but moderate bandwidths, leading to post-compressed pulses of a few 10's of fs and transmission efficiencies well above 95%. The second stage requires large bandwidth optics at the expense of lower reflectivity and damage threshold, yielding few-cycle pulses with efficiencies around 80%. This stage is the most critical since the shorter input pulses experience self-steepening (SS), dispersion and ionization, which can limit the spectral broadening and the output pulse compressibility. HCF-based post-compression has been very successful at reaching ultra-short pulse durations due to the shorter interaction lengths (dispersion free broadening) and the ability to partially compensate SS with ionization under specific conditions [3]. Some experiments even report hybrid nonlinear experiments involving a MPC as the first stage and a HCF for the second one [10]. Very recently, 290 µJ pulses from a Ti:Sa laser were post-compressed down to 3 optical cycles in a single-stage MPC with 45% efficiency [11]. This experiment is driven by sub-50 fs input pulses readily available from commercial Ti:Sa based laser systems, thereby avoiding the first MPC compression stage mentioned above.

In this paper, we report the post-compression of mJ-level 30 fs Ti:Sa laser pulses down to 5.3 fs (~2 optical cycles) with 67% overall efficiency. We spatially filter the input pulses in order to study MPC-based compression using near-perfect Ti:Sa beam properties. The spectral broadening capability is first explored at different gas pressures. The optimum configuration is then determined and the shortest accessible compressed pulse is characterized using single-shot second harmonic frequency-resolved optical gating (SHG-FROG). The intrinsic limitations and scalability of MPC-based compression of Ti:Sa pulses are discussed, providing reliable guidelines for further compression down to single-cycle duration.

The experimental setup is schematically shown in fig.1. It starts with a Ti:Sa based femtosecond laser system (Femtopower Pro-HE) delivering 30 fs compressed pulses of 1.2 mJ energy at 1 kHz repetition rate. An acousto-optic programable dispersive filter (Dazzler from Fastlite) in the main amplifier allows to finely tune the pulse spectral phase. The amplified beam is then spatially filtered under vacuum by a 21 cm long HCF of 250 µm inner diameter, which fixes the beam waist position and dimensions at its output. Its throughput of about 80% leads to pulses of 1 mJ energy.

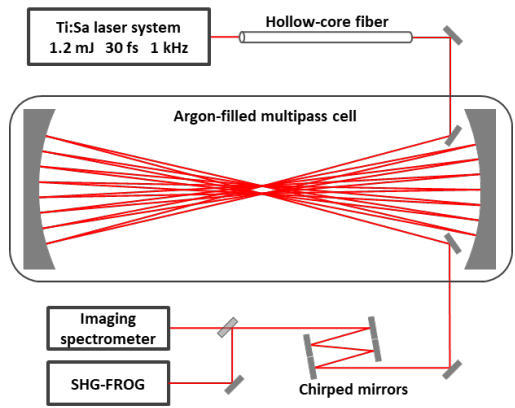

Fig. 1. Experimental setup

This HCF-based filtering stage is followed by a MPC made of two 2" enhanced silver-coated mirrors of 1.5 m radius of curvature separated by ~ 3 m, yielding a nearly concentric cavity. This geometry is set such that the beam size on the mirrors leads to a fluence well below their damage threshold, roughly estimated at 100 mJ/cm². The number of passes through the MPC can be set up to 18, and two small rectangular silver mirrors allow to inject/pick up the beam. The MPC is operated inside an airtight chamber filled with argon gas at different pressures.

A set of lenses after the hollow-core fiber matches the beam waist at the center of the MPC to a diameter of ~ 250 µm, which keeps the beam caustic periodic throughout the passes in the MPC. The Dazzler is tuned to compensate for dispersion (lenses, windows, air) such that the pulses are Fourier-transform limited (FTL) at the input of the MPC. The output beam is then temporally compressed by a set of double-angle chirped mirrors (PC 42 from Ultrafast Innovations) and two fused silica wedges to finely compensate the pulse spectral phase. The beam is then characterized spatially and temporally with the use of a single shot SHG-FROG and an imaging spectrometer (both devices from FemtoEasy).

The output spectrum is first studied at different gas pressures and for 18 passes through the MPC. Fig. 2 shows the spectral shapes obtained for pressures ranging from 0 to 700 mbar, and their respective RMS width and FTL pulse durations are plotted in Fig. 3. It demonstrates that the spectral broadening increases linearly with the gases pressure up to about 150 mbar but saturates at higher pressures. Indeed, the gas dispersion is also pressure dependent and plays a crucial role in the broadening process. At lower pressures, the propagation can be considered as dispersion free.

Here, all 18 passes are subject to pure SPM with a constant temporal pulse profile. The output spectra grow linearly with gas pressure and maintain a nice symmetry.

When reaching 150 mbar, the increasing spectral width along with the increasing dispersion coefficient of the gas start to temporally spread the pulses for the last MPC passes, thereby decreasing their peak-power and limiting spectral broadening. However, SPM along with dispersion lead to much smoother spectral profiles compared to dispersion-less spectral broadening like e.g. in HCF, where strong modulations appear with pure SPM.

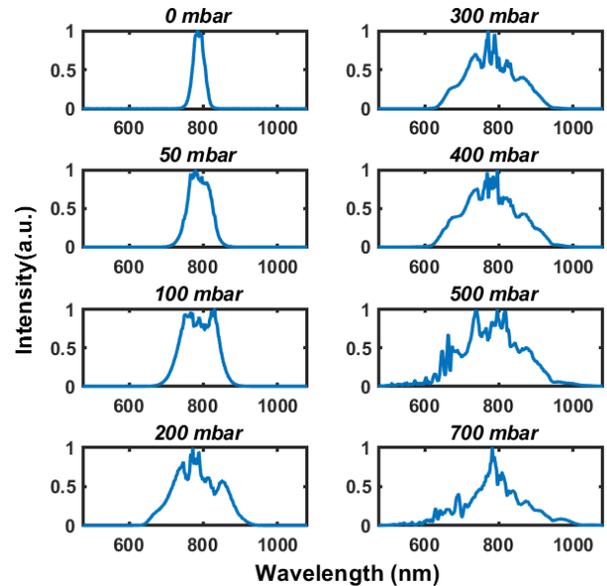

Fig. 2. Evolution of the normalized spectral shapes with increasing gas pressure.

At higher pressures, this effect is aggravated: although SPM becomes stronger, the pulse spreading arises earlier, at earlier passes, such that the next ones have little contribution to the spectral broadening. Besides, SPM accumulated by chirped pulses favor the onset of nonlinear spectral phase at high orders that can degrade the pulse compressibility. Thus, many passes at high pressure do not necessarily lead to shorter post-compressed pulses.

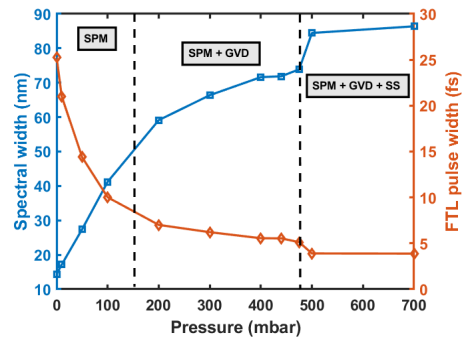

Fig. 3. Evolution of the RMS spectral widths and FTL pulse widths for different compression regimes.

In addition, starting from 475 mbar, the spectral profile exhibits noticeable distortions and a modulated pedestal towards shorter wavelengths. The onset of such distortions is very steep when increasing the gas pressure and shows a tipping point in the pulse spectral quality. However, this specific point does not correspond to the critical power above which the beam experiences catastrophic collapse, and the ionization rate of the gas remains sufficiently low to have a noticeable impact under these conditions.

Numerical simulations based on these experimental conditions show the exact same behavior and highlight an excessive impact of SS that arises during the first pass. This leads to an optical shock exhibiting a highly distorted pulse profile. The on-axis pulse temporal profile after the first pass is simulated and plotted in Fig. 5 for different gas pressures around the tipping point mentioned above. The optical shock of the pulse along with fast ripples at its trailing edge arise quite rapidly at this nonlinearity level. The whole on-axis pulse propagation over the first pass simulated for these specific gas pressures is presented in Visualization 1-3.

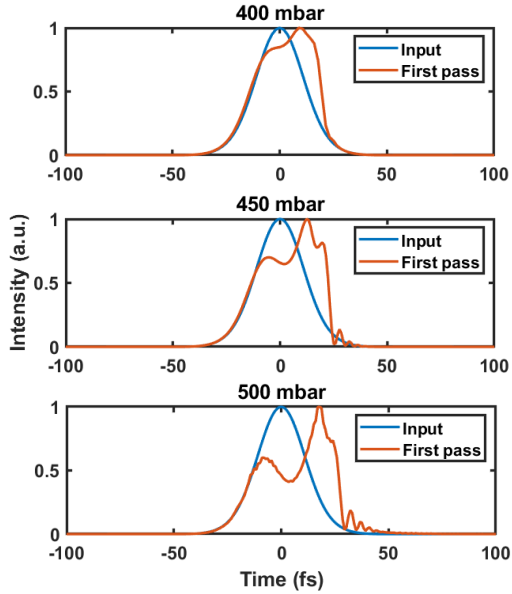

Fig. 4. Simulated on-axis pulse temporal profile for increasing gas pressure ranging from 400 mbar and 500 mbar.

The interplay between SPM, SS and the gas dispersion in the following passes induces strong temporal and spectral modulation in the short-wavelength region, which highly impacts the output spectral quality and pulse compressibility. It is worth noting that this effect cannot be avoided by e.g. limiting the number of passes since the first one is the most critical. Moreover, even for higher pressures, the spectral bandwidth remains almost constant. This tipping point fixes the fundamental limit of spectral broadening in the MPC regardless of the number of passes and directly depends on the dimensionless parameter $s$ characterizing the SS process ($s = 1/t_0\omega_0$, with $\omega_0$ the central pulsation and $t_0$ the RMS FTL pulse width) and the beam-averaged B-integral accumulated in the first pass. In our experimental conditions, they correspond to $s$ = 0.02 and $B$ = 5 rad.

This behavior is specific to short input pulses of few 10's of fs duration, where SS can arise quite rapidly with increasing SPM. Conversely, in the range of 100fs - 1ps, the B-integral per pass in the MPC is usually limited to avoid strong spatio-spectral couplings, but SS and gas dispersion have little or no impact. Besides, in HCF experiments for post-compressing pulses down to the single cycle regime, SS arises rapidly at low B-integrals, but can be partially compensated by ionization, which has the opposite effect for moderate ionization rates [3]. However, this situation relies on very specific experimental conditions and is usually avoided in MPCs due to plasma defocusing that disturbs the beam caustic stability throughout the passes.

This study demonstrates that single-stage MPC-based post-compression is limited in B-integral per pass by SS, and in number of efficient passes by dispersion. The gas pressure and the number of passes need to be chosen carefully, depending on the input pulse energy and duration. In the following, the number of passes is set to 16 and the gas pressure to 450 mbar, as a tradeoff between spectral broadening, pulse compressibility and MPC transmission efficiency.

In this configuration, the output energy of the MPC is 0.77 mJ compared to 0.98 mJ at its input, yielding a transmission of about 79%. This efficiency can be compared to the numbers of bounces on the silver mirrors, leading to a reflectivity of 98.5% for each pass over the whole spectral bandwidth. Out of the MPC, 9 pairs of chirped mirrors introducing a group-delay dispersion of 378 fs$^2$, followed by two wedges, allow to compensate for the accumulated nonlinear spectral phase and gas dispersion. The compressed energy is measured to be 0.66 mJ, including the beam transport losses, yielding 67% overall efficiency for the full post-compression system.

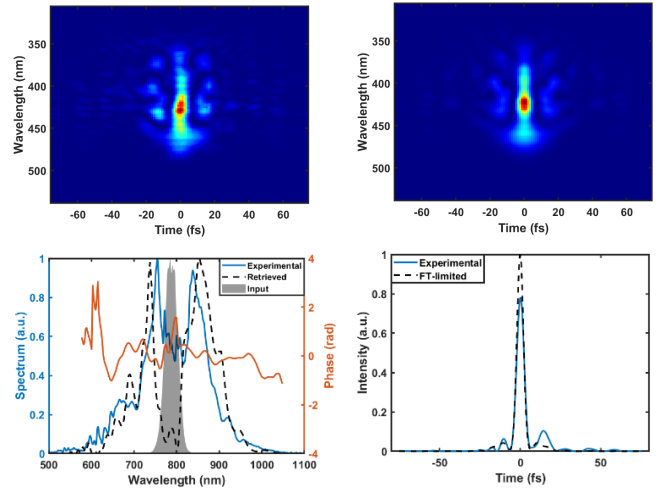

Fig. 5. Top: SHG-FROG traces measures (left) and retrieved (right). Bottom: input, measured and FROG-retrieved output spectra (left) and pulse temporal profile along with its FTL profile (right).

The output spectrum measured by a spectrometer is shown in Fig. 5 along with the one retrieved from the SHG-FROG measurement (error of 11.10$^{-3}$ on a 256x256 grid). The corresponding compressed pulse profile is presented in Fig. 3 and exhibits a duration of 5.3 fs FWHM, to be compared to its FTL duration of 4.7 fs. This yields a compression factor of about 5.7 and output pulses of two optical cycles.

This configuration is located close to the tipping point described above. Higher gas pressures and/or a higher number of passes lead to worse pulse compressibility and longer output pulse duration.

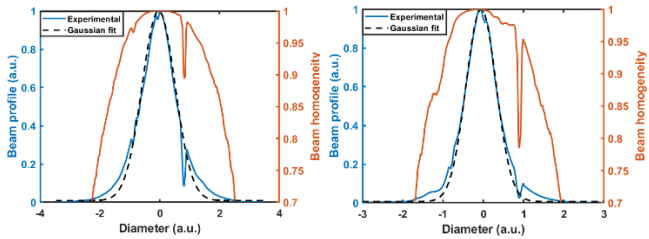

Fig. 6. Output beam profile in arbitrary units in the horizontal (left) and vertical (right) dimensions., along with their spectral homogeneity. The dip in the measurements is due to unremovable dusts on the spectrometer's slit.

The spectral homogeneity of the beam is characterized using an imaging spectrometer. The V parameter, as defined in [5], is plotted in Fig. 6 along with the beam profile along the horizontal and vertical dimensions. Note that the hole in the beam profile is caused by unremovable dusts on the spectrometer's slit. The output beam exhibits a nearly-Gaussian profile with very little degradation and nice spectral homogeneity, above 99% at the beam FWHM and above 95% at $1/e^2$ in both directions.

We demonstrated the generation of 5.3 fs pulses at 0.66 mJ output energy with excellent beam quality and homogeneity using single-stage nonlinear post-compression of 30 fs Ti:Sa pulses in an argon-filled MPC. This pulse duration corresponds to 2 optical cycles within the pulse FWHM. The transmission efficiency is measured to be 79% for the MPC itself and 67% for the overall compression experiment.

The spectral evolution of the pulses with respect to gas pressure was studied and highlighted two main limitations. The first one comes from the gas dispersion that progressively reduces the accumulated B-integral per pass in the MPC. However, it helps to maintain a nice spectral shape without significant modulations typically obtained in dispersion-less compression experiments. Nevertheless, the number of passes needs to be restricted to generate the largest spectrum with best temporal compressibility and limited MPC losses. On the other hand, the B-integral accumulated in the first pass needs to be carefully controlled to avoid dramatic temporal pulse distortions arising through SS. This situation cannot be exploited as the output pulse quality and compressibility are strongly deteriorated.

In our experimental conditions, these limitations restrict the compressed pulse to 2 optical cycles. Thus, the single-cycle regime would require careful dispersion management within the cavity to partially (but not fully) compensate for the gas dispersion. Indeed, a completely dispersion-less post-compression scheme would lead to a significant SS and highly unbalanced output spectral shapes. Tailored dispersion throughout the propagation would limit this effect while preserving smooth and weakly modulated spectral shapes.

Furthermore, spatial filtering of the input beam prior to the MPC, although implemented for practical reasons, in particular for beam matching inside the cavity, is not an absolute necessity for achieving compression. Our next experiments will involve direct output beams at higher energies and adapted cavity designs and compared to regular HCF post-compression experiments in terms of spectral and temporal pulse quality, overall post-compression efficiency and footprint in the single-cycle regime.

**Funding.** This project has received funding from the European Union's Horizon 2020 research and innovation program under grant agreement no. 871124 Laserlab-Europe and no. 694596 Advanced Grant ExCoMet, and from the Agence Nationale de la Recherche (ANR) under grant agreement ANR-10-LABX-0039-PALM.

**Disclosures.** The authors declare no conflicts of interest.

**Data availability.** Data underlying the results presented in this paper are not publicly available at this time but may be obtained from the authors upon reasonable request.

**References**

1. Tamas Nagy, Peter Simon & Laszlo Veisz (2021), Advances in Physics: X, 6:1.
2. Samuel Bohman, Akira Suda, Tsuneto Kanai, Shigeru Yamaguchi, and Katsumi Midorikawa, Opt. Lett. 35, 1887-1889 (2010).
3. Ouillé, M., Vernier, A., Böhle, F. et al., Light Sci Appl 9, 47 (2020).
4. Marc Hanna, Xavier Délen, Loic Lavenu, Florent Guichard, Yoann Zaouter, Frédéric Druon, and Patrick Georges, J. Opt. Soc. Am. B 34, 1340-1347 (2017).
5. J. Weitenberg, A. Vernaleken, J. Schulte, A. Ozawa, T. Sartorius, V. Pervak, H.-D. Hoffmann, T. Udem, P. Russbüldt, and T. W. Hänsch, Opt. Express **25**, 20502-20510 (2017).
6. L. Lavenu, M. Natile, F. Guichard, Y. Zaouter, X. Delen, M. Hanna, E. Mottay, and P. Georges, Opt. Lett. 43, 2252-2255 (2018).
7. Martin Kaumanns, Vladimir Pervak, Dmitrii Kormin, Vyacheslav Leshchenko, Alexander Kessel, Moritz Ueffing, Yu Chen, and Thomas Nubbemeyer, Opt. Lett. 43, 5877-5880 (2018)
8. Prannay Balla, Ammar Bin Wahid, Ivan Sytcevich, Chen Guo, Anne-Lise Viotti, Laura Silletti, Andrea Cartella, Skirmantas Alisauskas, Hamed Tavakol, Uwe Grosse-Wortmann, Arthur Schönberg, Marcus Seidel, Andrea Trabattoni, Bastian Manschwetus, Tino Lang, Francesca Calegari, Arnaud Couairon, Anne L'Huillier, Cord L. Arnold, Ingmar Hartl, and Christoph M. Heyl, Opt. Lett. 45, 2572-2575 (2020)
9. M. Müller, J. Buldt, H. Stark, C. Grebing, and J. Limpert, Opt. Lett. 46, 2678-2681 (2021).
10. L. Lavenu, M. Natile, F. Guichard, X. Délen, M. Hanna, Y. Zaouter, and P. Georges, Opt. Express 27, 1958-1967 (2019)
11. P. Rueda, F. Videla, T. Witting, G. A. Torchia, and F. J. Furch, Opt. Express 29, 27004-27013 (2021)